\documentclass[preprint,amsmath,amssymb,aps,prab,longbibliography,superscriptaddress]{revtex4-1}
\usepackage{graphicx}
\usepackage[caption=false,captionskip=0pt]{subfig}
\usepackage{listings}
\usepackage{lineno,hyperref}
\usepackage{color}
\usepackage{soul}
\usepackage{xcolor}

\modulolinenumbers[5]

\begin{document}

\title{Optimized computation of tight focusing of short pulses using mapping to periodic space}

\author{Elena~Panova}
\affiliation{Department of Mathematical Software and Supercomputing Technologies, Lobachevsky State University of Nizhni Novgorod, Nizhny Novgorod 603950, Russia}
\author{Valentin~Volokitin}
\affiliation{Mathematical Center, Lobachevsky State University of Nizhni Novgorod, Nizhny Novgorod 603950, Russia}
\affiliation{Department of Mathematical Software and Supercomputing Technologies, Lobachevsky State University of Nizhni Novgorod, Nizhny Novgorod 603950, Russia}
\author{Evgeny~Efimenko}
\affiliation{Institute of Applied Physics, Russian Academy of Sciences, Nizhny Novgorod 603950, Russia}
\author{Julien~Ferri}
\author{Thomas~Blackburn}
\author{Mattias~Marklund}
\affiliation{Department of Physics, University of Gothenburg, 41296 Gothenburg, Sweden}
\author{Alexander Muschet}
\author{Aitor De Andres Gonzalez}
\author{Peter Fischer}
\author{Laszlo~Veisz}
\affiliation{Department of Physics, Ume\aa \hspace{0mm} University, 90187, Ume\aa, Sweden}
\author{Iosif~Meyerov}
\affiliation{Mathematical Center, Lobachevsky State University of Nizhni Novgorod, Nizhny Novgorod 603950, Russia}
\affiliation{Department of Mathematical Software and Supercomputing Technologies, Lobachevsky State University of Nizhni Novgorod, Nizhny Novgorod 603950, Russia}
\author{Arkady~Gonoskov}
\affiliation{Department of Physics, University of Gothenburg, 41296 Gothenburg, Sweden}
\affiliation{Department of Mathematical Software and Supercomputing Technologies, Lobachevsky State University of Nizhni Novgorod, Nizhny Novgorod 603950, Russia}

\date{\today}

\begin{abstract}
When a pulsed, few-cycle electromagnetic wave is focused by {optics} with $f$-number
smaller than two, the frequency components it contains are focused to different regions of space, building up a complex electromagnetic field structure.
Accurate numerical computation of this structure is essential for many applications such as the analysis, diagnostics, and control of high-intensity laser-matter interactions.
However, straightforward use of finite-difference methods can impose unacceptably high demands on computational resources, owing to the necessity of resolving far-field and near-field zones at sufficiently high resolution to overcome numerical dispersion effects. Here, we present a procedure for fast computation of tight focusing by mapping a spherically curved far-field region to periodic space, where the field can be advanced by a dispersion-free spectral solver. In many cases of interest, the {mapping} 
reduces both run time and memory requirements by a factor of order 10, making it possible to carry out simulations on a desktop machine or a single node of a supercomputer. We provide an open-source C++ implementation with Python bindings and demonstrate its use for a desktop machine, where the routine provides the opportunity to use the resolution sufficient for handling the pulses with spectra spanning over several octaves. The described approach can facilitate the stability analysis of theoretical proposals, the studies based on statistical inferences, as well as the overall development and analysis of experiments with tightly-focused short laser pulses.
\end{abstract}

\maketitle



\section{Introduction}

There are well-established routines for computing the evolution of an electromagnetic field in many areas of computational physics.
Nevertheless, in some applications, the specific properties of the problem allow for tailored approaches that can reduce computational costs.
Tight focusing of short electromagnetic pulses is one of such problems.
{Tight focusing} is of growing interest in many research areas, including attosecond physics \cite{krausz.rmp.2009}, high-intensity laser-plasma interactions \cite{mourou.rmp.2006, marklund.rmp.2006}, as well as laser-based studies of fundamental quantum systems \cite{dipiazza.rmp.2012}. Apart from being central for reaching high field strengths for a given radiation power \cite{yanovsky.oe.2008}, the use of tight focusing of few-cycle pulses with specific phase- and polarization properties enables opportunities for controlling laser-matter interaction scenarios \cite{chatziathanasiou.p.2017, harvey.prl.2017}. 

{While analytical models of a tightly focused electromagnetic field (e.g., paraxial~\cite{davis.pra.1979, barton.jap.1989, sheppard.josa.1999, sepke.ol.2006, salamin.apb.2006}} or beyond paraxial~\cite{couture.pra.1981, narozhny.jetp.2000, lin.prl.2006, fedotov.spie.2007, sapozhnikov.ap.2012}) are widely used in many theoretical studies, numerical computation provides a direct approach that is of particular interest, as it avoids intrinsic inaccuracies associated with the underlying assumptions \cite{yu.spie.2019} and is more flexible in terms of initial conditions.
Paraxial models, in particular, require that the diffraction angle be small and therefore become increasingly inaccurate as the focusing becomes stronger.
Numerical computations can therefore be necessary when tight focusing is used to achieve strong fields at a target, and the interaction process is sensitive to the field structure in the focal region. This is the case in many situations, such as vacuum electron\linebreak acceleration \cite{popov.pop.2008, bochkarev.qe.2007} and the generation of radiation using laser-driven individual \cite{harvey.prab.2016} or collective \cite{gonoskov.pre.2011, kormin.ncom.2018, cardenas.scirep.2019} dynamics of electrons~\cite{popov.pop.2008, bochkarev.qe.2007,harvey.prab.2016,gonoskov.pre.2011, kormin.ncom.2018, cardenas.scirep.2019}. Numerical computations can be also crucial when the laser pulse is only few cycles long \cite{rivas.scirep.2017} and, therefore, cannot be treated analytically as monochromatic.

Besides the use in simulations \cite{thiele.jcp.2016, perez.pre.2019}, numerical computations can also support experimental efforts to achieve high intensity, e.g., by controlling adaptive optics \cite{bahk.apb.2005} or by retrieving information from the measured output based on the solution of inverse problem\cite{gonoskov.scirep.2019}. Numerical computations are also of clear interest for designing experiments \cite{gonoskov.prl.2013, gonoskov.prl.2014, gefler.pra.2015, gonoskov.prx.2017, vranic.ppcf.2017, gong.pre.2017, efimenko.scirep.2018, efimenko.pre.2019, magnusson.prl.2019, magnusson.pra.2019} at the next generation large-scale laser facilities \cite{danson.hplse.2019}, where strong electromagnetic fields are likely to be reached by the combination of several, tightly focused laser pulses~\cite{bulanov.prl.2010, gonoskov.pra.2012}. 

Tight focusing of few-cycle pulses is an emerging topic with a broad field of applications \cite{naumova.pop.2005}. Recently, the few-cycle laser technology has reached the state to produce higher energies much beyond the mJ level, which in combination with tight focusing provides strongly relativistic peak intensities \cite{rivas.scirep.2017} for various promising sources of isolated attosecond XUV pulses \cite{kormin.ncom.2018} and relativistic electron bunches \cite{cardenas.scirep.2019}. However, this tight focusing is still very challenging \cite{yanovsky.oe.2008, bahk.apb.2005}, especially with the almost octave spanning spectrum of few cycle laser pulses, and it will profit from further support by quick numerical algorithms.

The numerical computation of electromagnetic field in the focal region for a given field in far-field zone can be performed via direct integration using the Green's function. In the most general case, when the phase and/or polarization properties vary across the transverse direction of the pulse to be focused, the computational costs are proportional to $N M$, where $N$ and $M$ are the number{s} of nodes of the grids that sample the field in the far- and near-field zones respectively. Note that the lowest spatial frequency components of the pulse determine the distance to and, thus, the size of the far-field zone, whereas the highest spatial frequency components determine the necessary resolution. This leads to an extensive growth of computational costs in the case of tight focusing (i.e., large diffraction angle) and short pulse duration, which is of interest in the context of the problems discussed above. The use of finite-difference time-domain (FDTD) methods for evolving the field from the far-field to the near-field zone leads to computational costs that are proportional to \emph{NT}{$\sim$}
\emph{N}$^{4/3}$, where \emph{T}{$\sim$}
\emph{N}$^{1/3}$ is the number of time steps to be performed. Such computations are subject to numerical dispersion, which cannot be removed along all the directions simultaneously and may affect significantly high-frequency components of the pulse~\cite{blinne.cpc.2018}.

Another attractive option is to use a spectral field solver, since it is free of numerical dispersion and it is possible to advance the field over large time intervals in a single step, leading to the numerical costs proportional to $N \ln N$, assuming the use of a fast Fourier transform (FFT). In this case, however, the computations can still be hampered by the extensive size of the grid used for sampling the pulse in the far-field zone. The use of supercomputers, on the other hand, is complicated by the nonlocal data processing patterns required by the FFT routine.

In this article, we present a computational scheme that makes use of the fact that a short pulse in the far-field zone is located within a thin spherical layer and thus occupies only a small fraction of the entire simulation domain. Taking advantage of the periodicity of spectral solvers, we map the far-field zone onto a smaller grid, achieving a significant reduction of computational costs. The implementation of the spectral solver and the proposed scheme was performed as part of the hi-$\chi$ (Hi-Chi, High-Intensity Collisions and Interactions) project \cite{hichi} (see also \cite{muraviev.arxiv.2020}). This open-source software package is a collection of Python-controlled tools for performing simulations and data analysis in the research area of strong-field particle and plasma physics. All the components are being developed in C++ and optimized for high performance CPUs, which allows for efficient performance of resource-intensive calculations. As part of the Hi-Chi project, we have implemented an easy-to-use and high-performance tool for modeling tightly focused radiation that can be used for computations on desktops (the supercomputer version is under development). In this work we validate the developed computational module and use it to characterize the properties of focused radiation, namely, the dependency of peak amplitude and the location of the attained peak on the $f$-number.

{The presented method and the developed routine can be of interest for both theoretical and experimental studies that involve tightly focused pulses, especially if the pulses have few-cycle durations. We would like to outline three specific potential applications {that involve the calculation of the tight focus a large number of times}: (1) {stability/robustness analysis of the outcome of theoretical proposals 
with respect to spatial} laser-pulse imperfections; (2) {iterative determination of the input parameters (beam profile and wavefront) from other experimental observables, e.g., electron spatial distribution in a specific experiment}, which can even be performed on many single-shot measurements;
(3) generation of large sets of simulations for training machine learning methods employed for diagnosis and improvement of tight focusing by means of adaptive optics.}

\section{Materials and Methods}

\subsection{Spectral Solvers}
\label{sec:application_of_spectral_solvers}

An essential part of the scheme we present is the use of a spectral (Fourier) method to advance the electromagnetic field in a region with periodic boundary conditions. In this subsection we briefly outline the main equations and practical aspects of this approach.

The central idea of spectral methods for time-dependent problems is to make use of the fact that, for linear homogeneous equations with a well-determined nondegenerate spectrum, the state can be represented via eigenstates that can be independently advanced over an arbitrary large time interval without computational errors. When eigenstates correspond to plane waves, the transformation from the representation on a spatial grid to the eigenbasis representation (and vice versa) can be done with a FFT. Source terms (i.e., inhomogeneous part) and/or nonlinear terms can be accounted for by the step-splitting technique (see, for example, \cite{weideman.jna.1986, birdsall.1991}). Since the propagators for these terms do not necessarily commute with that for the homogeneous equation, this technique introduces errors that grow with, and thus restrict, the size of the time step. Nevertheless, since we consider the propagation of electromagnetic radiation in vacuum, the source term (governed by the current) is zero and we thus can do arbitrarily large time steps.

The application of split-step spectral method to the Maxwell equations has been carried out and used in many numerical studies (see, for example, \cite{haber.cnsp.1973, lin.pf.1974, buneman.jcp.1980, gustafsson.1995, vay.jcp.2013, gonoskov.phd.2013} or chapter 15.9~(b) in \cite{birdsall.1991}). 
We denote the electromagnetic field and current density in Fourier space by $\hat{\textbf{E}}$, $\hat{\textbf{B}}$ and $\hat{\textbf{j}}$ respectively.
{Furthermore,} $\hat{\textbf{k}}$ is the normalized wave vector $\textbf{k}$, $i$ is the imaginary unit, $c$ is the speed of light, {and} $C=\cos(ck\Delta t)$, and $S=\sin(ck\Delta t)$. Using superscripts $n$ and $n+1$ to denote consecutive numerical states, we can express the Fourier method in the following form (see \cite{vay.jcp.2013}, CGS units are used): %
\begin{align}
\label{eq:PSATD_E}
\begin{split}
\hat{\textbf{E}}^{n+1}=C\hat{\textbf{E}}^{n}+\textit{i}S\hat{\textbf{k}}\times\hat{\textbf{B}}^{n}-\frac{4\pi S}{kc} \hat{\textbf{j}}^{n+1/2}&+(1-C)\hat{\textbf{k}}(\hat{\textbf{k}}\cdot\hat{\textbf{E}}^{n}) \\
&+4\pi\hat{\textbf{k}}\left(\hat{\textbf{k}}\cdot\hat{\textbf{j}}^{n+1/2}\right) \left(\frac{S}{kc}-\Delta t\right), \\
\end{split} \\
\label{eq:PSATD_B}
\begin{split}
\hat{\textbf{B}}^{n+1}=C\hat{\textbf{B}}^{n}-\textit{i}S\hat{\textbf{k}}\times\hat{\textbf{E}}^{n}+4\pi\textit{i}\frac{1-C}{kc}\hat{\textbf{k}}\times\hat{\textbf{j}}^{n+1/2}. \\
\end{split}
\end{align} 

It is worth noting that Poisson's equation $\hat{\textbf{k}}\cdot\hat{\textbf{E}}=0$ is not satisfied here automatically but maintained approximately. This may lead to the appearance of static electric fields that arise from a spurious charge associated with either the initial field state or accumulated computational errors (in case of computations with nonzero current). {To correct for this, we need to subtract the longitudinal component of electric field $\hat{\textbf{k}}(\hat{\textbf{k}}\cdot\hat{\textbf{E}})$ from the initial electric field $\hat{\textbf{E}}$ once, before the simulation starts (for more details see \cite{vay.jcp.2013} or Section 6.2 in~\cite{gonoskov.phd.2013}). This will ensure that Poisson's equation is satisfied throughout the simulation and that the corresponding term in Equation~(\ref{eq:PSATD_E}) vanishes. However, to make our solver capable of handling continuous sources, and to avoid the accumulation of numerical errors due to imperfect satisfaction of the continuity equation in nonvacuum simulations (beyond the scope of this article), we carry out this adjustment at each iteration:}
\begin{eqnarray}
\label{eq:PSATD_E_pois}
&&\hat{\textbf{E}}^{n+1}=C(\hat{\textbf{E}}^{n}-\hat{\textbf{k}}(\hat{\textbf{k}}\cdot\hat{\textbf{E}}^{n}))+\textit{i}S\hat{\textbf{k}}\times\hat{\textbf{B}}^{n} \\
\label{eq:PSATD_B_pois}
&&\hat{\textbf{B}}^{n+1}=C\hat{\textbf{B}}^{n}-\textit{i}S\hat{\textbf{k}}\times\hat{\textbf{E}}^{n}.
\end{eqnarray} 

As our implementation follows that referred to as PSATD in Ref.~\cite{vay.jcp.2013}, we adopt the same terminology for our spectral solver.

\subsection{Problem Statement}
\label{sec:problem}

We assume that, at the start of the simulation, the pulse that is to be advanced to focus is located within a spherical sector, defined by a polar angle $\theta$ around the negative $x$ axis that can be as large as $\pi/2$. The opening angle $\theta$ is related to the $f$-number, $f$, as $\theta = \arctan\left(1/(2f)\right)$.

The distance to the far-field zone is determined by the lowest characteristic spatial frequency component of the pulse. Since we are interested in the treatment of few-cycle pulses, we can take the central wavelength $\lambda$ (the one that corresponds to the mean of the frequency band) as the typical scale for the lowest spatial frequency component.
{Then we require} that the distance $R_0$ from the initial location of the pulse to the geometrical center (focus position) is much larger than $\lambda$. In practice we set $R_0 \approx 16 \lambda$ but this can vary depending on the spectral bandwidth and on the required level of accuracy.

The pulse can have arbitrary profiles along the longitudinal and transverse directions, i.e., perpendicular to, and along, the spherical surface respectively.
The pulse can have arbitrary variation of {profile}, polarization and phase (wavefront) along the transverse directions, but the variations should not be too rapid, so that we still can assume that locally the pulse propagates predominantly towards the geometrical center of the sphere. The permitted magnitude of variations depends on the particular problem and on the required accuracy level. This is a matter of numerical verification.

Although the problem statement is not restricted to any {particular} form, we consider a specific case of a short pulse under tight focusing. This facilitates the description of the method and provides a good demonstration of its capabilities.
To mimic a laser beam being uniformly amplified within {the amplifying} 
medium, we consider a pulse with flat-top transverse profile. In this case we need to define the transverse shape $u_{ts}$ so that the pulse amplitude is nonzero only within the angle $\theta$. In order to provide a smooth transition near this angular limit, we introduce an edge smoothing angle $\varepsilon$ and use $\cos^2$ shape:
\begin{eqnarray}
\label{eq:uts2}
u_{ts}(\alpha)=\omega\left(\alpha, 0, \theta-\frac{\varepsilon}{2}\right) + \left( \cos{\frac{\pi\left(\alpha-\theta+\frac{\varepsilon}{2}\right)}{2\varepsilon}} \right)^2 \cdot \omega\left(\alpha, \theta-\frac{\varepsilon}{2}, \theta+\frac{\varepsilon}{2} \right),
\end{eqnarray}
where we use the rectangular function
\begin{eqnarray}
\label{eq:window}
\omega \left(x, x_{min}, x_{max}\right)=
\begin{cases}
1~\text{for}~x \in \left[x_{min}, x_{max}\right] \\
0~\text{for}~x \notin \left[x_{min}, x_{max}\right]
\end{cases}.
\end{eqnarray}

We consider a longitudinal shape of the form:
\begin{eqnarray}
\label{eq:ul}
u_l(x)=\sin \left(\frac{2\pi x}{\lambda}\right) \cdot \cos^2\left(\frac{\pi x}{L}\right) \cdot \omega\left(x, -\frac{L}{2}, \frac{L}{2}\right),
\end{eqnarray}
where $L$ is the pulse length, $\alpha$ is the angle between the $x$ axis and the position vector $\textbf{R}=(x,y,z)$.

Finally, the spherical pulse can be defined as
\begin{eqnarray}
\label{eq:u}
&&u\left(x,y,z\right){=u\left(\textbf{R}\right)}=\frac{A}{R} \cdot u_l\left(R-R_0\right) \cdot u_{ts}\left(\arcsin\frac{\sqrt{y^2+z^2}}{R}\right) \\
\label{eq:A}
&&A=\sqrt{\frac{4P_0}{c(1-\cos{\theta})}},
\end{eqnarray}
where $P_0$ is the input power. We consider linear polarization and set the electromagnetic field as follows:
\begin{eqnarray}
\label{eq:E}
&&\textbf{E} (\textbf{R}) = u(\textbf{R}) \bf s_0 \\
\label{eq:B}
&&\textbf{B} (\textbf{R}) = u(\textbf{R}) \bf s_1,
\end{eqnarray}
where $\bf s_0$ and $\bf s_1$ are, respectively, the normalized vectors ${\textbf{d}}\times{\textbf{d}}\times\textbf{R}$ and ${\textbf{d}}\times\textbf{R}${,} 
and ${\textbf{d}}$ is the polarization vector.

\begin{figure}[h]
\includegraphics[width=0.7\linewidth]{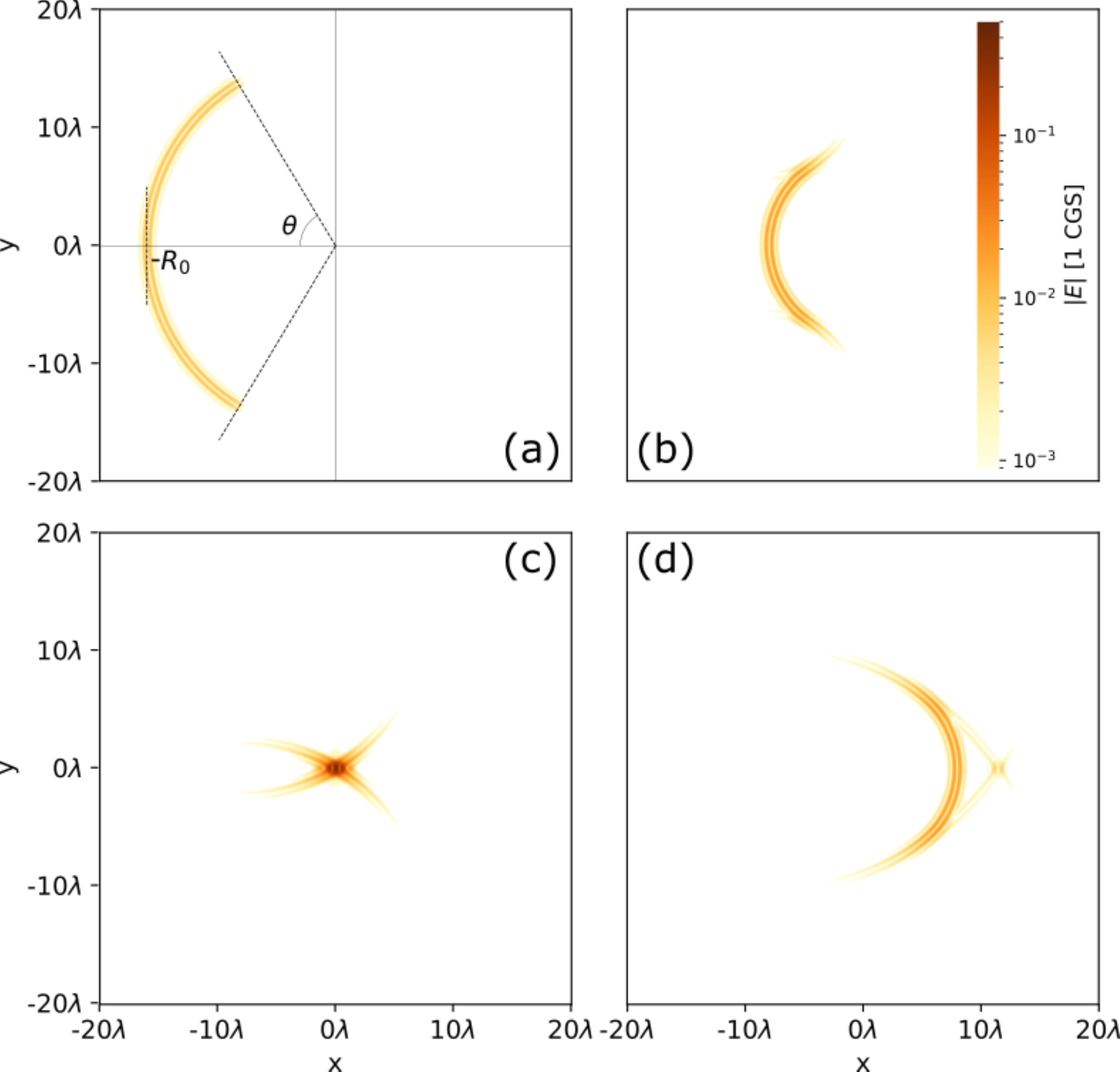}
\caption{An example of simulation of the tight focusing problem. The electric field intensity is shown for (\textbf{a}) $t = -R_0/c$, (\textbf{b}) $t = -R_0/2c$, (\textbf{c}) $t = 0$ and (\textbf{d}) $t = R_0/2c$ with f{$-$}number$=0.3$ ($\theta \approx 1$~rad), $\lambda=1$~$\mu$m, $R_0=16\lambda$, $L=2\lambda$, $\varepsilon=0.1$~rad, $P_0=1$~W. The pulse propagates towards positive $x$ direction, the transverse directions are spanned by $y$ and $z$ axes (see Equation~(\ref{eq:u})).}
\label{fig:full_calc}
\end{figure}

An example of simulation of the tight focusing problem by the method described in Section~\ref{sec:application_of_spectral_solvers} is demonstrated in Figure~\ref{fig:full_calc}.

\subsection{Mapping to and from the Computational Subregion}
\label{sec:simulation_in_thin_layer}

The stated problem of advancing electromagnetic field can be solved numerically by sampling and iterating the electromagnetic field in the region $\left(x \in \left(-R_0 - L/2, L/2\right)\right) \cap\left(y \in \left(-R_0, R_0\right)\right) \cap\left(z \in \left(-R_0, R_0\right)\right)$, which has volume equal to approximately $4 R_0^3$ (assuming that $R_0 \gg L$). However, geometric arguments indicate that the pulse should remain, approximately, in a thin spherically curved layer $R_0-L/2-ct < R < R_0+L/2-ct$. Thus only a small fraction of the computational grid actually contains the pulse. If we were using {the Finite Difference Time Domain (FDTD)} method, we could reduce computational costs by updating the grid nodes in the nonempty regions only. However, as we wish to avoid numerical dispersion, we use a Fourier-based solver and thus do not have such a straightforward opportunity. We now describe how we can effectively exclude the empty regions and achieve a significant reduction of computational costs in this case.

\begin{figure}[h]
\includegraphics[width=0.7\linewidth]{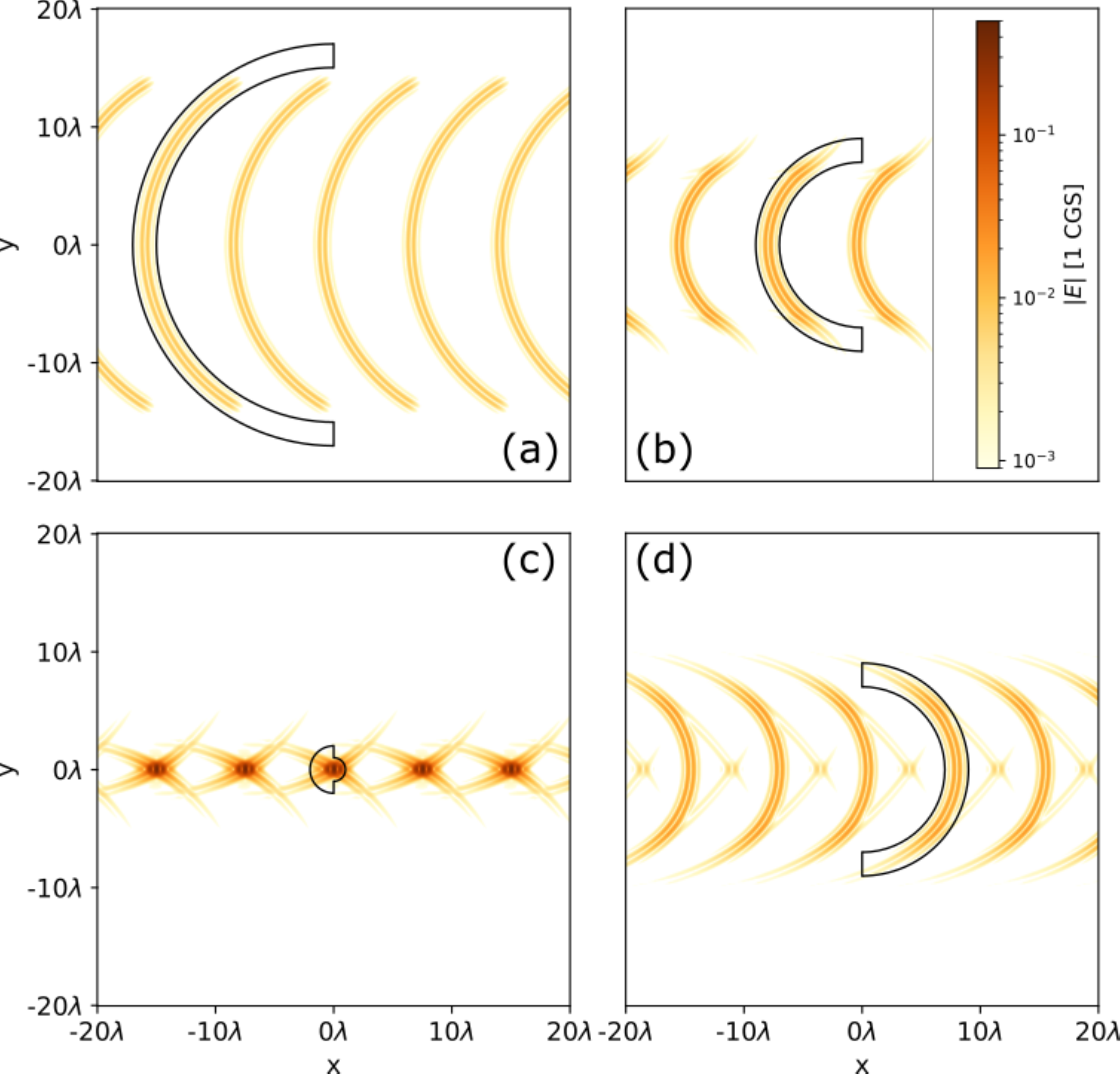}
\caption{Illustration of the idea of the proposed method. 
 The pictures show the evolution of the electromagnetic field composed by the equidistant replication of the initial pulse along the $x$ axis: the field intensity is shown for (\textbf{a}) $t = -R_0/c$, (\textbf{b}) $t = -R_0/2c$, (\textbf{c}) $t = 0$ and (\textbf{d}) $t = R_0/2c$ with f{$-$}number$=0.3$ ($\theta \approx 1$~rad), $\lambda=1$~$\mu$m, $R_0=16\lambda$, $L=2\lambda$, $\varepsilon=0.1$~rad, $P_0=1$~W.}
\label{fig:reverse}
\end{figure}

Let us consider a periodic field structure containing a series of repetitions of the pulse, spaced apart by a distance $D$ along the $x$ axis (see Figure~\ref{fig:reverse}a). If $D$ is large enough, the pulses overlap neither initially nor during their propagation towards the focal region (see Figure~\ref{fig:reverse}b). When each pulse reaches {its} 
focus, a sequence of identical structures of focused pulses is formed (see Figure~\ref{fig:reverse}c). These structures are spaced apart by the distance $D$ and have only a minor overlap due to diffractive spreading of the signal from the fringes of the initial structure. We can simulate this process by computing the evolution of the field in a region that has periodic boundary conditions and size $D$ along $x$ axis. The volume of this region is $4 D R_0^2 \ll 4 R_0^3$ (we assume that $D \sim L \ll R_0$). Periodic boundary conditions are effectively provided by the Fourier solver by default. In this way we can obtain the numerical solution of the stated problem at reduced computational cost, with respect to both run time and memory. Let us describe particular details of our implementation.

\begin{figure}[h]
\includegraphics[width=0.9\linewidth]{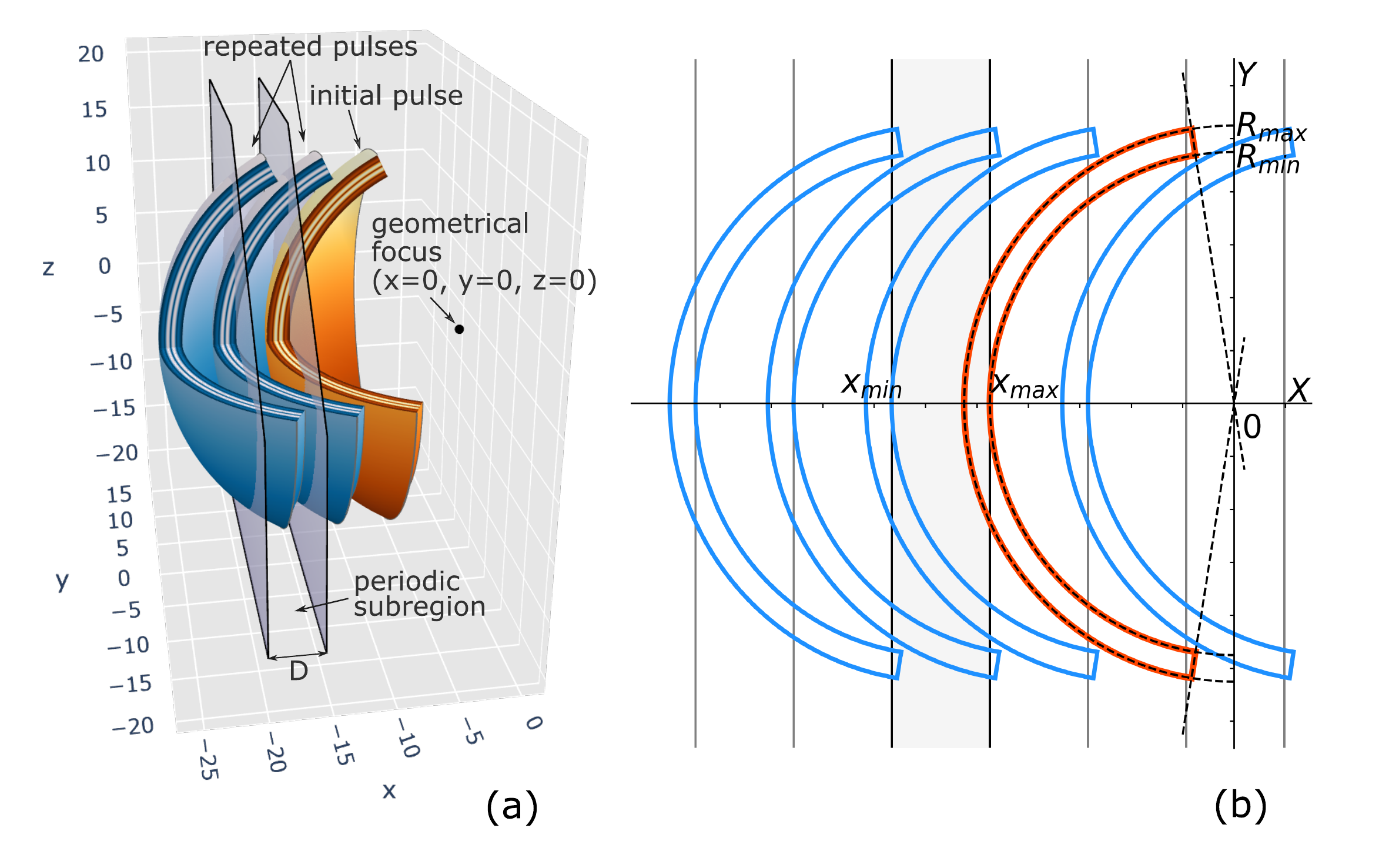}
\caption{Schematic illustration of the proposed method in 3D space (\textbf{a}) and the cross-section in the $x$-$y$ plane at $z = 0$ (\textbf{b}). The restriction given by Equation~(\ref{eq:D}) is derived from requiring that the initial (red) and replicated (blue) pulses do not overlap. The shaded region is the subregion where the simulation is performed.}
\label{fig:pulses_3d_image}
\end{figure}

The choice of parameter $D$ can be restricted by requiring that neighboring pulses do not overlap. From the geometrical consideration shown in Figure~\ref{fig:pulses_3d_image}, one can obtain the following restriction:
\begin{eqnarray}
D \geq D_{min} = - R_{min}\cos{\theta}+\sqrt{R^2_{max}-R^2_{min}\sin^2{\theta}},
\label{eq:D}
\end{eqnarray}
where $R_{max} = R_0 + L/2$ and $R_{min} = R_0- L/2$ are the boundaries of the initial pulse in spherical coordinates. Note that one can avoid this restriction by considering a superposition of the overlapping replicated pulses. In this case we can still obtain the separated structure of a single pulse {at the focus} 
if $D > L$ due to the linearity of the Maxwell's equations. However, due to overlap, in this case we are not able to obtain (in a straightforward way) the intermediate states of the pulse on the way to the focus. The choice of $D$ can also depend on the extent of fringes and the required accuracy. There is a trade-off{: increasing $D$ reduces} 
the errors but increases the computational costs of the method.

Although this choice of periodic domain is arbitrary with respect to translation along $x$, it is useful to describe a particular implementation in detail. We suggest to carry out calculations in a layer $x \in \left[-R_{min}-D, -R_{min}\right]$ with periodic boundary conditions. In order to set the initial conditions we need to perform the following mapping {$\textbf{T}\left(\textbf{R} \right)$}: the field state of each point $(\tilde{x}, \tilde{y}, \tilde{z})$ of the simulated subregion is assigned to that of point $\left(x, y, z\right)$ of the original unlimited space, for which the problem is stated: 
\begin{eqnarray}
\left(\tilde{x}, \tilde{y}, \tilde{z}\right)= \textbf{T}\left(\textbf{R} \right) = \left( x_{min} +D\,\left\{\frac{x-x_{min}}{D}\right\}, y, z \right).
\label{eq:mapping}
\end{eqnarray}

Here $\left\{\cdot\right\}$ is the fractional part of an argument.

Unfortunately, there is no reverse mapping. However, by considering the geometrical propagation of the pulse, we can ``cut out'' the primary pulse from the periodic space. We do so in the following way: for each point $\textbf{R}$ of the original space the field state $\textbf{u} = \left(\textbf{E}, \textbf{B} \right)$ is assigned using the field state $\tilde{\textbf{u}}\left(\textbf{T}\left(\textbf{R}\right)\right)$ within the simulated region according to:
\begin{align}
\textbf{u}(\textbf{R}, t)=\tilde{\textbf{u}}(\textbf{T}\left(\textbf{R}\right))\cdot I_{S(t)}(\textbf{R})
\label{eq:rev_mapping_u}
\end{align}
\begin{align}
I_{S(t)}(\textbf{R})=
\begin{cases}
0, & \left(r(t)<0\right)~{\text{and}}~\left(\left(R\geq-l(t)\right)~{\text{or}}~\left(R<-r(t)\right)~{\text{or}}~(x>0)\right) \\
0, & \left(l(t)\leq 0\right)~{\text{and}}~\left(r(t)\geq 0\right)~{\text{and}}~\left(x<0\right)~{\text{and}}~\left(R>L\right) \\
0, & \left(l(t)\leq 0\right)~{\text{and}}~\left(r(t)\geq 0\right)~{\text{and}}~\left(x\geq 0\right)~{\text{and}}~\left(R>l(t)\right) \\
0, & \left(l(t)>0\right)~{\text{and}}~\left(\left(R\leq l(t)\right)~{\text{or}}~\left(R>r(t)\right)~{\text{or}}~(x<0)\right) \\
1, & \text{in other cases} \\
\end{cases}
\label{eq:rev_mapping_I}
\end{align}
\begin{align}
l(t)=-R_{max}+ct,~r(t)=-R_{min}+ct,~\textbf{R}=(x,y,z),~R=||\textbf{R}{||}
\label{eq:rev_mapping_lr}
\end{align}
The described method for cutting out a single pulse from the periodic sequence is shown in Figure~\ref{fig:reverse} using black contours.

This method does not always provide an ideal cut-off of secondary pulses, which can still fall into the region $S(t)$ denoted by {Equations~(\ref{eq:rev_mapping_I}) and (\ref{eq:rev_mapping_lr})}, if $D$ is not large enough. However, in practice, it is usually sufficient to take the parameter $D$ equal to $3L-4L$ in order to avoid notable presence of the secondary pulses. However, depending on the shape of the transverse component of the spherical wave, the fringes may show up in the obtained results and indicate the necessity of choosing larger value of $D$. The choice of the parameter $D$ is described in more detail in Section~\ref{sec:verification}.

We would like to conclude this section by making a few general remarks. Firstly, the choice of $R_0$ in many cases is determined by the transition from the geometrical to wave optics. If the deviations from the spherical shape of the phase front are sufficiently minor, one can combine the proposed method with the computation of the earlier field evolution within geometrical optics. This implies advancing the pulse along the rays pointing to the geometrical center in combination with the adjustment of the field amplitude inverse proportionally to the distance to the center. Note that the effect of imperfect phase front can be also estimated and taken into consideration. Anyway, in many cases it should be possible to compute the pulse at the entrance to the near-field zone, which our method is designed for. Secondly, our method makes use of the solver periodicity along the main direction of propagation ($x$ axis) and effectively reduces the computation in a {cube} to the computation in a thin layer that spans to a large distance along the transverse directions ($y$~and $z$ axes) only. Using this as a basis, one can further {duplicate} the pulse several times along the transverse directions, making use of the solver periodicity in these directions and of the linearity of the Maxwell's equations. In the considered case the effective speedup scales as $R_0$, whereas the described follow-up development would yield a speedup that scales as $R_0^3$.

Finally, we estimate the expected benefit of using the method. According to the discussion above, the method reduces both computational time and memory demands by a factor of $R_0/D$ (neglecting the logarithmic factor in the computational complexity of the FFT routine). While the choice of $R_0$ depends on the required level of accuracy, an estimate appropriate in many cases is 10 times the wavelength of the radiation. For long pulses this would mean $R_0 \sim 10 \lambda$. However, since the method is designed for very short pulses (with duration equivalent to a few cycles) {we assume an octave-spanning spectrum and}
account for the shape of the pulses {by using}
the longest wavelength component
$L$, i.e., {$R_0\sim 10L$}.
As we will see further (see Section~\ref{sec:verification}) a relative error in the peak field amplitude of less than 1\% is reached at $D \approx 2L$. Thus we estimate that the described method can reduce the computational time and memory demands by a factor of order 10 (Section~\ref{sec:verification}), depending on the level of required accuracy.
\\
\\
\\
\\
\\

\section{Results and Discussion}

\subsection{Verification and Accuracy Determination}\label{sec:verification}

The mapping described in Section~\ref{sec:simulation_in_thin_layer} allows us to reduce the time and memory costs significantly. However, the question arises as to how accurate the computational results are. As compared to the direct computation, the only possible inaccuracy is the one caused by the overlap of the pulse in the periodic space. The accuracy therefore can be controlled by the ratio $D/L$, which defines the relative size of margins used for avoiding overlaps. For a given required accuracy, the acceptable value of $D/L$ depends on the problem parameters, primarily on the f-number and on the sharpness of the fringes mentioned above. To provide a general idea about the achievable accuracy and the corresponding values of $D/L$, we compare the results of the described method to the results obtained via direct computation.

In order to verify and examine the capabilities of the method in the most extreme conditions we take the field configuration described in Section~\ref{sec:problem} and consider a large value of the opening angle $\theta = 1$ ($\text{f-number}\approx 0.3$) and short pulse duration $L = 2\lambda$. We take the edge smoothing angle $\varepsilon = 0.1$. We assume that the pulse is focused along the $x$-axis and is initially located within the spherical layer in the region $x<0$ at a sufficiently large distance $R_0 = 16\lambda$ from the origin. For this case the direct computation can be performed in the region $x,y,z\in[-20\lambda, 20\lambda]$. The proposed method can be applied using the region with reduced size $D$ along the $x$-axis. 
Because of the large value of $\theta$, the peak intensity is reached almost exactly at the geometrical focus, i.e., at time $t = R_0/c$ {since the start of the simulation}. Due to the properties of the spectral solver (Section~\ref{sec:application_of_spectral_solvers}), we can do only one iteration of the method with a time step of $\Delta t=R_0/c$ to obtain the electromagnetic field at the instant of focusing. For the simulation, a computational grid with the resolution of 48 points per wavelength along $x$-axis and 1536 $\times$ 1536 points along $y$- and $z$-axis ($y,z\in[-20\lambda, 20\lambda]$) was used. The error was calculated as the maximal difference of the field measured between the cases of direct computation and the computation with the considered method.

The computations for $D = 7{\lambda} = 3.5L$ show that the error introduced by the presented method is 0.12\% relative to the peak amplitude obtained in the direct computation. The time required for the computation is reduced by a factor of almost 6. The memory costs for the presented method were 55 GB RAM, instead of 300 GB RAM required for the direct~computation. 

\begin{figure}[h]
\includegraphics[width=1.0\linewidth]{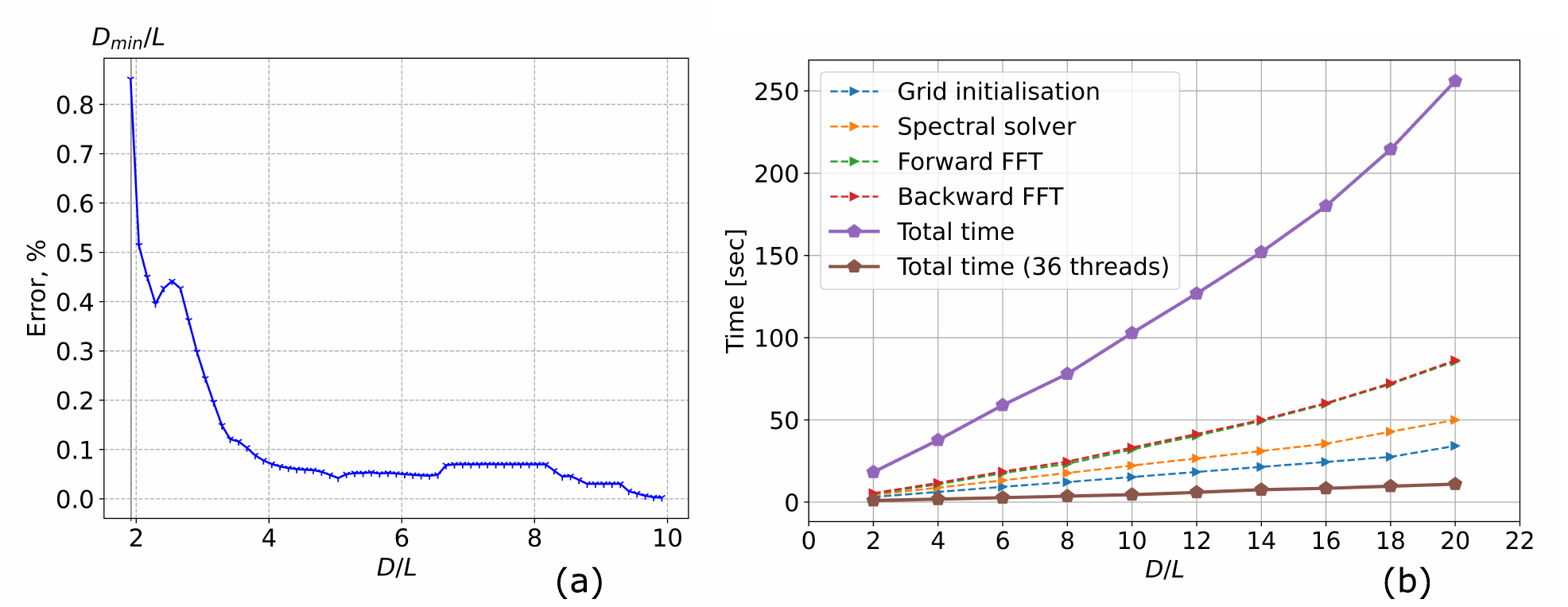}
\caption{(\textbf{a}) Relative error of the peak amplitude, comparing the computations in a subregion to that in the full domain, as a function of $D$, the subregion size. Here $\text{f-number}=0.3$. {(\textbf{b}) Computational time of one iteration of the method (grid initialisation, forward FFT, spectral solver execution, and backward FFT), as a function of the parameter $D/L$. All the simulations were performed using Hi-Chi. Computation time of sequential (all lines except the brown one) and multithreaded (brown line, 36 physical cores) versions of the code are presented. Size of the computational grid varies with the parameter $D$ and is equal to $(56 D/L)\times$ 896 $\times$ 896.}}
\label{fig:error_D}
\end{figure}

The dependence of the relative error on $D$ for a grid with the resolution of 12 points per wavelength along $x$ axis is shown in Figure~\ref{fig:error_D}a. As one can see, even at $D_{min}/L \approx 1.9$ the error does not exceed 0.9\%, and it quickly falls below 0.1\% at $D/L \approx 3.5$. Figure~\ref{fig:reverse} shows that the small error values at $D/L > 3.5$ can be attributed to the diffraction of the fringes that do not focus with the remainder of the pulse.

{All of the results in this and further sections were obtained with the open source Hi-Chi framework \cite{hichi}. The computation time of a typical simulation for different values of parameter $D$ is presented in Figure~\ref{fig:error_D}b. The experiments were performed on the MVS-10P supercomputer at the Joint Supercomputing Center of the Russian Academy of Sciences. Each computational node of the supercomputer consists of 2$\times$18-core Intel Xeon Gold 6154 CPU (3.0 GHz) and 192 GB of RAM. The code was built with Intel C++ Compiler 17.0.4, and the high-performance multithreaded FFT implementation from the Intel Math Kernel Library 2017 (MKL) was employed. For double precision, we find that the computation time and memory cost per grid cell are 230 ns (in single-thread mode) and 48~bytes, respectively. Therefore, while the simulation in entire region ($D/L = 20$) for a grid with the resolution of 28 points per wavelength takes 256~s and 40 GB of memory, the simulation using the presented scheme with $D/L = 2$ takes only 18 s and 4 GB of memory. Thus, we are able to reduce the calculation time and memory costs by a factor of 10, while the error introduced by the method did not exceed 0.85\%.}

\subsection{Examples}\label{sec:results}

In this section we provide three illustrative examples. In the first example we compute focusing of Gaussian beams for different values of f-number and compare the focused beam field with that obtained analytically within the paraxial approximation. This example both provides independent validation based on the known analytical result and indicates the importance of using the complete computation for small values of f-number. In the second example we again consider the described case of a short pulse with a circular flat-top transverse profile and compute the peak field strength as a function of f-number. The obtained data provides an opportunity to assess the potential benefit of using tight focusing for increasing peak field strength at current and future laser systems. Finally, in the third example, we apply the method to illustrate the degradation of the intensity profile at focus when using real laser pulses. This shows the necessity to describe accurately the laser pulse for obtaining realistic computational predictions and for interpreting the experimental data.

\subsubsection{Focusing of a Gaussian Laser Pulse}

In this subsection we compare our simulation results with theoretical predictions for a focused Gaussian beam. In this case, a simulation is initialized with a linearly polarized, spherical pulse with transverse electric field $E_\perp = E_0 R_0 \sin[k(r - R_0)] u_l(r - R_0) u_{ts}(\theta) / r$, where the longitudinal profile $u_l(\zeta) = \exp[-\zeta^2 \ln 4 / (n \lambda)^2]$ and the transverse profile $u_{ts}(\theta) = \exp(-\theta^2 / \theta_0^2)$.
Here $n$ is the number of wavelengths corresponding to the pulse duration (full width at half maximum) and the diffraction angle $\theta_0$ is related to the f-number $f$ by $\tan\theta_0 = 1 / (2 f)$.
The amplitude $E_0$ is determined by requiring that the instantaneous power transmitted through the spherical surface $r = R_0$ (the pulse peak) is equal to a fixed value $P_0$, i.e., $E_0 = \sqrt{P_0 / [\pi \varepsilon_0 c R_0^2 F(\theta_0)]}$ where $F(\theta_0) = \int_0^\pi \sin(\theta) \exp(-2\theta^2/\theta_0^2) \,\mathrm{d}\theta$.
For $\theta_0 < 2$ (equivalent to $f > 0.2$), this auxiliary function is well approximated as\linebreak $F(\theta_0) \simeq \theta_0^2 / 4 - \theta_0^4 / 48 + \theta_0^6 / 960$.
We set the wavelength $\lambda = 1.0~$micron, the number of cycles $n = 4$, the initial position of the pulse $R_0 = 20\lambda$, the peak power $P_0 = 1$~W, and vary $f$ from $0.25$ to $2$.
The simulation covers the domain $-30\lambda < x < 10\lambda$ and $-20\lambda < y, z < 20\lambda$, which is represented by $512$ grid points in each dimension. 
    
\begin{figure}[h]
    \includegraphics[width=0.97\linewidth]{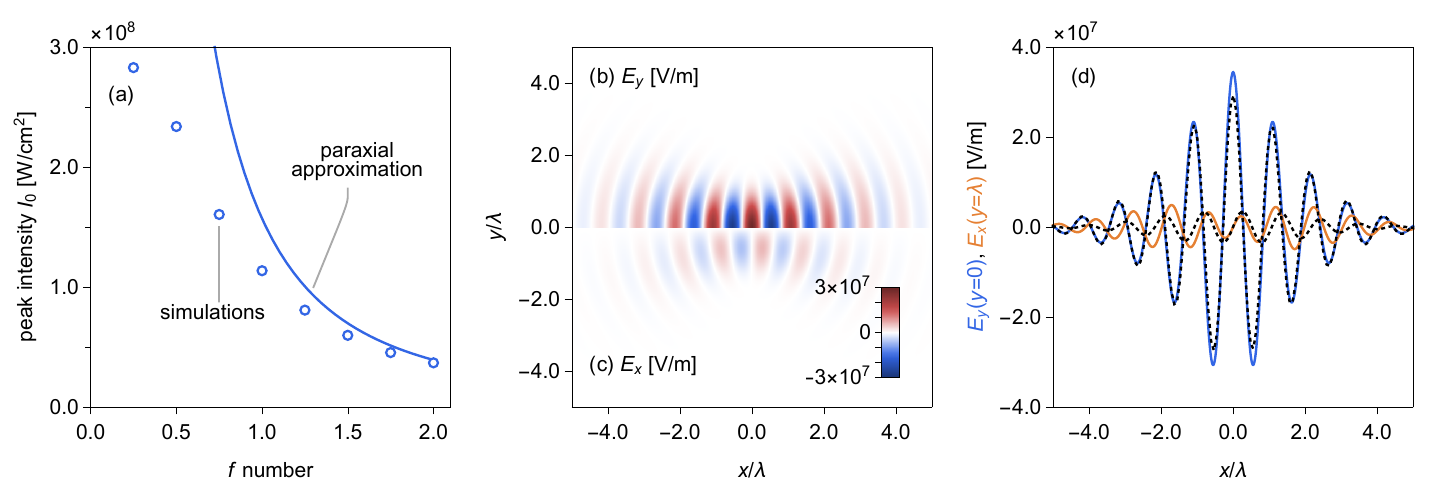}
    \caption{%
        Comparison of numerical and analytical results for a tightly focused pulse with Gaussian temporal and angular profiles and fixed peak power $P = 1$~W:
        (\textbf{a}) The peak intensity at focus.
        The (\textbf{b}) transverse and (\textbf{c}) longitudinal fields in the plane $z = 0$, at focus, from simulations with an $f$-number of 1.
        (\textbf{d}) Comparison of theoretical (solid colors) and simulation (black, dashed) results for the transverse field along $y = z = 0$ (blue) and the longitudinal field along $y = \lambda$ and $z = 0$, at focus, for $f = 1$.
    }
    \label{fig:ParaxialComparison}
    \end{figure}

The peak intensity at focus, as obtained from simulations, is shown in Figure~\ref{fig:ParaxialComparison}.
If $f < 1$, there are substantial deviations from the theoretical prediction for a paraxial Gaussian beam $I_0 = 2 P_0 / (\pi w_0^2)$, where $P_0$ is the instantaneous power passing through the focal plane $z = 0$ and the waist $w_0 = 2 f \lambda / \pi$ (see \cite{salamin.apb.2006}).
As the latter is calculated to lowest order in the diffraction angle, or for large $f$, this is to be expected.
A more detailed comparison for the specific case $f = 1$ is shown in Figure~\ref{fig:ParaxialComparison}b--d.
While the transverse field along the laser axis is reasonably well predicted by theory, the longitudinal component is not.
This supports the necessity of going beyond the paraxial approximation when considering tightly focused laser pulses.
However, we also confirm that our simulations correctly reproduce the expected theoretical result in the limit that the focusing becomes~weak.

\vspace{-12pt}
\subsubsection{Focusing of a Laser Pulse with a Circular Flat-Top Transverse Profile}

\begin{figure}[h]
\centering
\subfloat[]{
\includegraphics[width=0.435\linewidth]{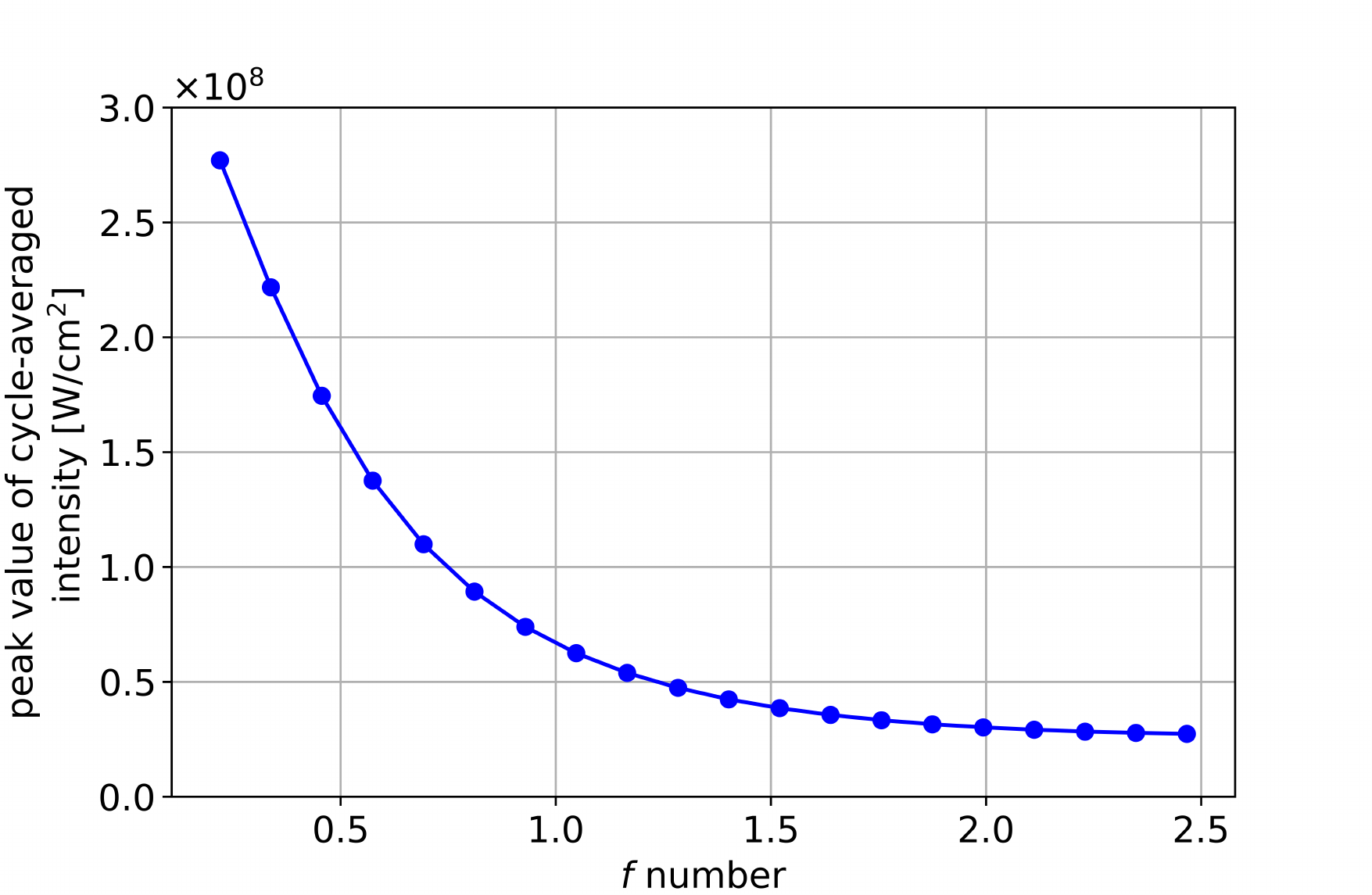}
\label{fig:peak}}
\subfloat[]{
\includegraphics[width=0.4\linewidth]{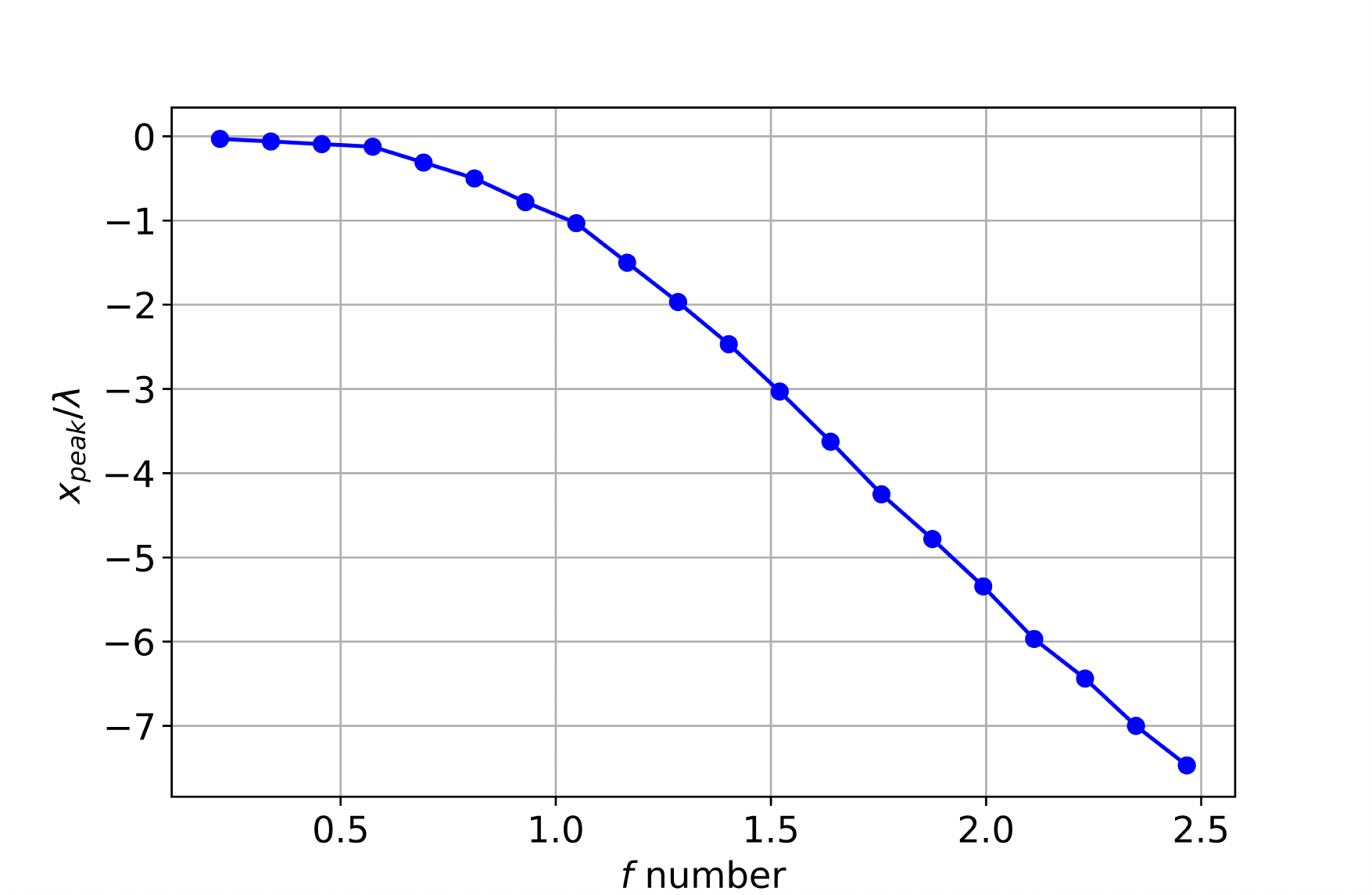}
\label{fig:x_peak}}
\caption{The dependence of (\textbf{a}) the peak value of cycle-averaged intensity on $\text{f-number}$ and (\textbf{b}) the coordinate $x$ at which the peak is reached, at a fixed peak power $P_0=1$~W and wavelength $\lambda=1$~$\mu$m (see the explanation in the text).}
\label{fig:peak_intensity}
\end{figure}

In this subsection we apply the developed method for computing tight focusing of laser beams with more realistic structure. To account for the uniform amplification within the gain medium and the importance of not exceeding the breakdown intensity threshold, we consider the flat-top model of the pulse described in Section~\ref{sec:problem} with the field configuration parameters presented in Section~\ref{sec:verification}. We then apply the described method to compute the peak field strength, as well as when and where it is reached. Figure~\ref{fig:peak_intensity} shows the dependency of the peak field amplitude and the coordinate at which it is reached on the f-number.
The distance $D$ between pulses was taken to be $2L$, so that the error of the scheme does not exceed 1\% of the peak amplitude (Section~\ref{sec:verification}). The size of the computational grid was 128 $\times$ 1024 $\times$ 1024 nodes, and the instant of reaching peak power was determined with an accuracy of $\Delta t=0.01${~s}. The complete code by which these results were obtained is given in the {Supplementary Materials}. 

Note that for f-number~$\lesssim 0.6$ the peak field strength is achieved almost at the geometrical center of the initial spherical phase front. Nevertheless, for larger values of f-number the place of reaching the strongest field moves away from the geometrical centre with increase of f-number. This is the result of competition between geometrical propagation and diffraction.
One can understand the nature of this mismatch through considering the limit of large $f$-numbers. When approaching to this limit, the pulse is limited by an increasingly small angle $\theta$ and the phase front is increasingly close to plane under this limitation. In this case the pulse almost immediately starts transverse spreading due to diffraction, i.e., the field amplitude starts to decrease. That is why the peak field is achieved far away from the geometrical center of the sphere that defines the phase front. (If we fix the transverse size of the pulse, the center itself tends to be infinitely far from the initial location of the pulse in the limit of large $f$-numbers.) From this explanation we see that this mismatch is originated from the diffraction of the pulse in the initial state. This means that for larger values of $f$-number we would need to take a larger value of $R_0$ to have the initial pulse in the zone of geometrical optics, i.e., so that it is well-described by a spherical wavefront. A simple estimate is to scale $R_0$ with the Rayleigh length, i.e., $R_0 \sim 10 \lambda f^2$. In brief, for large values of $f$-number, one should either account for the outlined shift (if approximate calculation is sufficient) or adjust the value of $R_0$ accordingly (preferred).

\vspace{-12pt}
\subsubsection{Focusing of Realistic Laser Pulses} 

\begin{figure}[]
    \includegraphics[width=\linewidth]{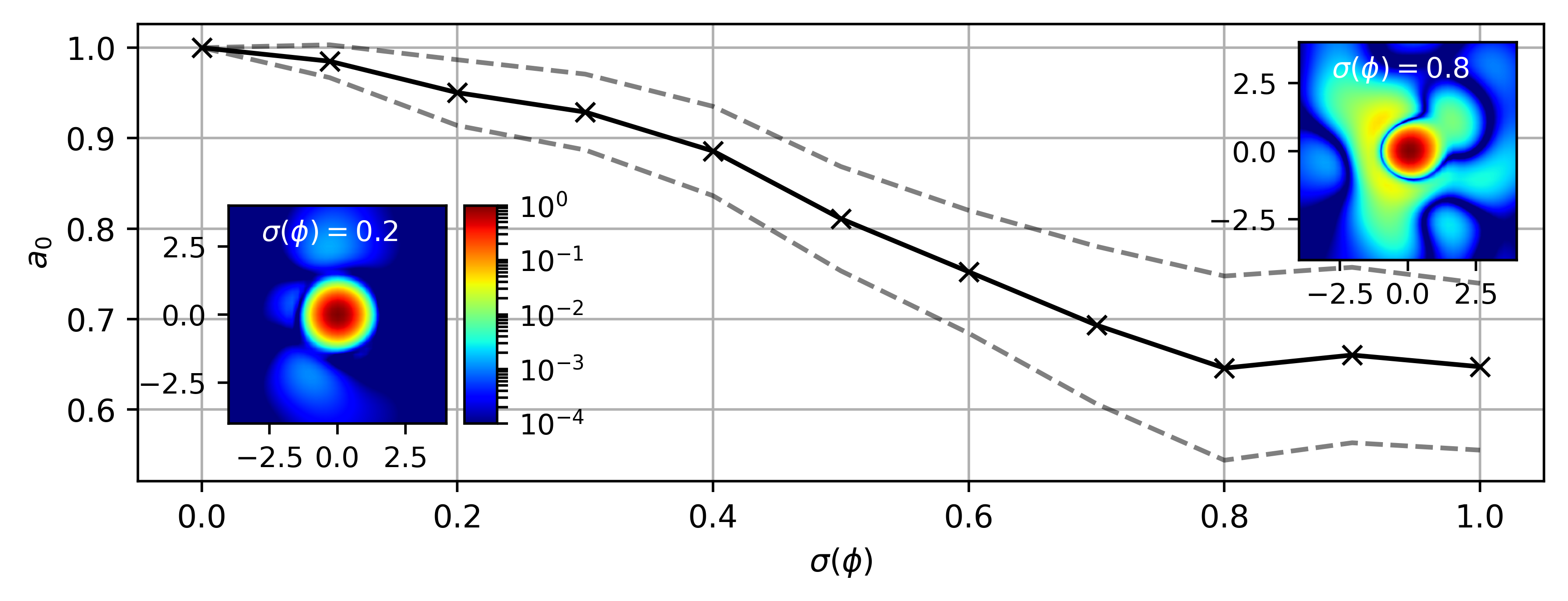}
    \caption{%
        Evolution of the peak amplitude $a_0$ at focus as a function of the standard deviation of the initial phase ($\sigma(\phi)$ in radian, solid line) and standard deviation of $a_0$ over 20 runs (dashed lines). Inlets exhibit examples of the normalized transverse intensity profile obtained at focus for two different values of $\sigma(\phi)$. Transverse directions are normalized to $\lambda$.}
    \label{fig:phase_pert}
    \end{figure}

We now modify the flat top profile used previously to introduce perturbation of the phase that could stand for the imperfections of an experimental wavefront, in the case of a f-number $f=1$.
The perturbation is described by a sum of Gaussians randomly located in the initial transverse profile, whose individual amplitude varies randomly between $\pm \varepsilon_\phi$, but with fixed width $\sigma=0.3\theta$, with $\theta$ the opening angle. The number of individual {perturbation} scales is $(2\theta/\sigma)^2=44$. Figure~\ref{fig:phase_pert} shows how the maximum intensity at focus varies with the standard deviation of the phase $\sigma(\phi)$. This clearly indicates that initial perturbations of the wavefront can significantly influence the intensity profile at focus, with a steady decrease of the maximum intensity as the perturbation increases. Thus, one has to give some thoughts to the quality of the laser system when designing experiments, as applications of tight focusing pulses are highly dependent on the maximum intensity.

\begin{figure}[h]
    \centering
    \begin{minipage}{.45\linewidth}
        \centering
        \includegraphics[width=1.\linewidth]{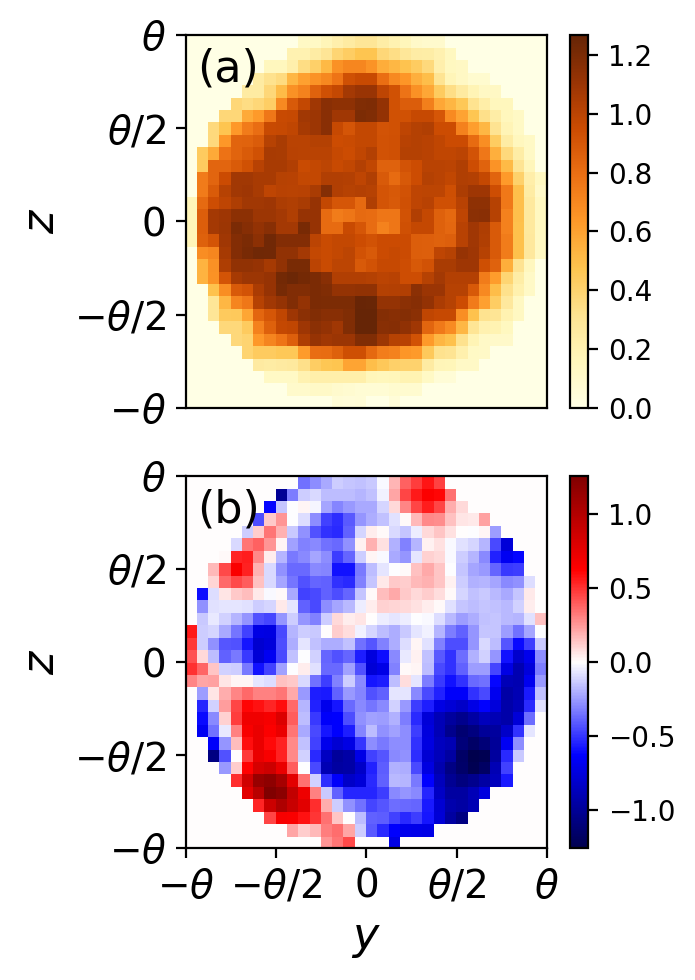}
    \end{minipage}%
    \begin{minipage}{0.55\linewidth}
        \centering
        \includegraphics[width=1.\linewidth]{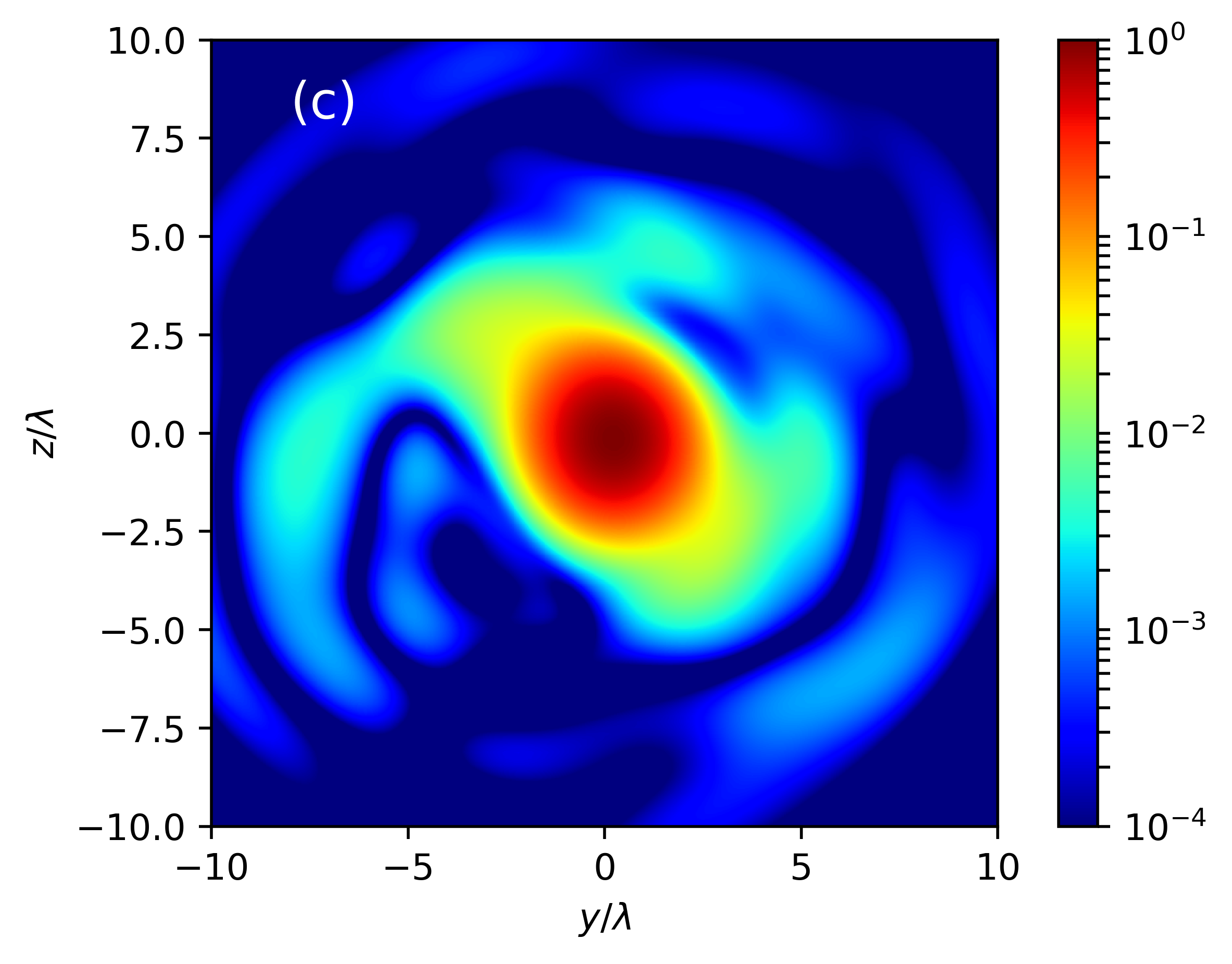}
    \end{minipage}
        \caption{%
        Example of focal spot calculated from realistic {profile}: 
        (\textbf{a}) typical transverse intensity {profile} 
        (a. u.) and (\textbf{b})  wavefront ({in} 
        radians) without correction of extra aberrations during beam transport from the end of the laser to the experiment and (\textbf{c}) normalized focal intensity obtained applying our method with a f-number $f=3$.}
    \label{fig:exp_prof}
\end{figure}

Another useful feature is that the described method allows for a fast computation of the laser focal spot given the initial beam profile and wavefront measured in an experiment. In Figure~\ref{fig:exp_prof}a the beam profile of the red spectral range (700--1000~nm) of the Light Wave Synthesizer 20 (LWS-20) at Ume\r{a} is plotted, while (b) shows the wavefront on the {focusing optics}
, which was originally corrected at the end of the laser and collected extra aberrations during transport in the vacuum compressor and beam line including about 10 mirrors. The calculated LWS-20 focal spot with $f=3$ without further corrections is highlighted in\linebreak Figure~\ref{fig:exp_prof}c. 
If applied directly in an experiment such a spot is applicable for simulation purposes. Note that other methods can also be used to retrieve beam profiles using experimental inputs, such as the Gerchberg--Saxton algorithm~\cite{gerchberg.optik.1972} which was, for example, used for laser wakefield simulations with realistic laser profiles~\cite{ferri.scirep.2016}. Alternatively, the experimental focus can be optimized with the help of our method coupled with machine learning techniques and using the measured beam profile before focusing and focal spot. This compensates for all added aberrations from the beam transport as well as the {focusing~optics}.

\vspace{-6pt}
\section{Conclusions}

We have presented an open-source implementation (see \cite{hichi}) of a method for optimized numerical computation of tight focusing of short electromagnetic pulses with arbitrary longitudinal and transverse variation of polarization, phase, and amplitude. The method is based on mapping 3D space to and from a periodic domain, which can have a much smaller size than that required for the straightforward computation. For example, for the case of a two-cycle pulse focused with f-number~$=0.3$, the computational costs for both run time and memory can be reduced by a factor of almost 6, while the introduced error with respect to the field strength does not exceed 0.1\% of the peak value.

We use a spectral solver to avoid numerical dispersion and to make it possible to advance the field state over an arbitrary time interval within one iteration. This provides an opportunity to perform computations of practical interest within minutes on a personal~computer.

The method and its implementation ha{ve} been validated by the comparison with the {straightforward computations (i.e., without using the described method)} and also with the results of analytical computations based on the paraxial approximation for Gaussian beams. We have also considered tight focusing of a laser pulse with a circular flat-top transverse profile. We apply the method to compute the peak achievable intensity as a function of f-number, indicating the potential benefit of using tight focusing for reaching high field strength with current and future laser facilities. Two realistic applications are also demonstrated including the focusing of an ideal and a real laser beam profile with imperfect wavefronts. These extend the potential utilization of our technique with performing simulations with real experimental data and an enhanced correction of experimental focus quality even for low repetition rate lasers that will have a high impact on future~experiments.



\vspace{6pt} 

\section{Data availability}
The Hi-Chi project is available online at \url{https://github.com/hi-chi/pyHiChi}. The scripts required to reproduce the numerical results are located in the ``doc/Tight Focusing'' folder.

\section{Acknowledgements}
The work is supported by the Lobachevsky University academic excellence program (Project 5-100). A.G. acknowledges the support of the Swedish Research Council (Grant No. 2017-05148), and L.V. also acknowledges the support of the Swedish Research Council (2016-05409 and 2019-02376). The authors acknowledge the use of computational resources provided by the Joint Supercomputer Center of the Russian Academy of Sciences and by the Swedish National Infrastructure for Computing (SNIC).




\end{document}